\pdfoutput = 1

\documentclass[11pt,twocolumn,a4paper]{article}

\usepackage{units}
 \usepackage[final]{graphicx}
\usepackage{rotating}
\usepackage{amsmath}
\usepackage{empheq}

\usepackage{rotating}
\usepackage{authblk}
\usepackage{color}
\usepackage{natbib}

\title{On the Electron Agyrotropy during Rapid Asymmetric Magnetic Island Coalescence in Presence of a Guide Field}

\author[1]{E.~Cazzola\thanks{emanuele.cazzola@wis.kuleuven.be}}
\author[1]{M.~E.~Innocenti}
\author[2]{M.~V.~Goldman}
\author[2]{D.~L.~Newman}
\author[3]{D.S.~Markidis}
\author[1]{G.~Lapenta}

\affil[1]{Department of Mathematics, KULeuven University, Celestijnenlaan 200B, Leuven, 3001, Belgium.}
\affil[2]{Center for Integrated Plasma Studies, University of Colorado Boulder,
Boulder, Colorado, USA.}
\affil[3]{PDC Center for high Performance Computing, KTH Royal Institute of Technology, Teknikringen 14, 10044 Stockholm, Sweden.}

\begin{document}

\maketitle
%
%

\begin{abstract}

We present an analysis of the properties of the electron velocity distribution during island coalescence 
in asymmetric reconnection
with and
without guide field. In a previous study, three main domains were identified,
in the case without guide field, as X-, D- and M-regions featuring
different reconnection evolutions \citep{cazzola2015}. These regions are also
identified here in the case with guide field. We study the departure
from isotropic and gyrotropic behavior by means of different robust detection algorithms proposed in the
literature. While in the case without guide field these metrics show an
overall agreement, when the guide field is present a discrepancy in the agyrotropy
within some relevant regions is observed, such as at the separatrices
and inside magnetic islands. Moreover, in light of the new
observations from the Multiscale MagnetoSpheric mission, an analysis of the electron
velocity phase-space in these domains is presented.

\end{abstract}

\section{Introduction}

Magnetic reconnection is a highly multi-scale physical process occurring in plasmas when magnetic field lines with opposite polarity come in contact.
The process releases a large amount of the stored magnetic energy after a complete restructuring of the magnetic field topology.
This effect makes reconnection one of the most important sources of accelerated particles in space.
However, magnetic reconnection alone cannot explain the high energy particles measured in some regions in space, and other accelerating processes have to be 
taken into account alongside reconnection.
One of the most solid explanation involves multiple acceleration mechanisms during the formation, growth and coalescence of magnetic islands \citep{drake2006, oka2010}.
While the effects of island coalesce in symmetric reconnection have been sufficiently studied over the decades, the same process in asymmetric configuration is still 
poorly investigated, especially when a
strong guide field is present.

In a previous work, \citet{cazzola2015} have shown that, during rapid island coalescence with no guide field, three different reconnection regions can be observed. 
These regions have been identified as X-regions, 
where similar traces as the traditional asymmetric X-point are observed,
D-regions, where reconnection occurs between two diverging islands and reveals an opposite behavior with respect the X-regions, 
and M-regions, where
the reconnection event occurs between two merging magnetic islands.
In this work, a similar analysis leads to the identification of these three types of regions also in the case with guide field.
It is important to identify parameters that allow high resolution satellites, such as the Magnetospheric MultiScale (MMS) NASA mission \citep{burch2014}, to 
distinguish between these three regions.

This work intends to give more insights into the electron behavior in these three specific regions, in support of future comparisons with observational data.
Particular attention will be given to the departure of particles from the initial 
isotropy and gyrotropy.
In Particle-In-Cell (PIC) simulations, the departure from gyrotropy is quantified from information collected in the pressure tensor.
Several methodologies are available.
Hereafter, we will refer to these methodologies as \emph{detection algorithms} to underline their numerical nature
based on a precise mathematical algorithm.
The algorithms used in this work are those proposed by \citet{scudder2008} ($A\O{}$), \citet{aunai2013} ($D_{ng}$, 
shown in the supporting material)
 and \citet{swisdak2015} ($\sqrt{Q}$). 
The first method focuses on particle agyrotropy in the plane perpendicular to the local magnetic field. It relies on the diagonalization of 
a pressure tensor specifically built upon the perpendicular velocity direction, and easily retrievable from traditional PIC pressure tensors
 after some mathematical manipulation.
The other two methods propose different algorithms which also include the parallel component
in the computation. 
The mathematical formulation of these three metrics is summarized in the suporting material. 
%
None of these methods has been  applied so far to the case of asymmetric island coalescence, either with or without guide field.

In addition to the previous quantities, we visualise electron agyrotropy 
by carrying out an analysis of the electron temperature in the frame of reference based on the local magnetic field. 
The parallel and the two orthogonal perpendicular directions to the local magnetic field $T_{||}$,  $T_{\perp_1}$ and $T_{\perp_2}$,  are identified
according to the definition given in \citet{goldman2015} and recalled in Section \ref{sec:simulation}.
The analysis of agyrotropy is addressed with the ratio $\nicefrac{T_{\perp_1}}{T_{\perp_2}}$, while the ratio $\nicefrac{T_{||}}{T_{\perp_1}}$ is used to study the anisotropy.
$\nicefrac{T_{\perp_1}}{T_{\perp_2}}$,  gives insight into \emph{relative departure} from agyrotropy in the two direction $\perp_1 \text{ and } \perp_2$, 
which are not necessarily the directions where the temperature 
differs more.
Notice the latter approach is strictly simulation-frame dependent, whereas
quantities $A\O$, $Q$ and $D_{ng}$ instead address the perpendicular components independently of the simulation frame adopted. These quantities are computed from simulations results only, by not directly involving the 
simulation axis.
A systematic comparison between the aforementioned metrics will be given to remark on their principal differences. 
Since the method proposed in \citet{aunai2013} is conceptually similar to \citet{swisdak2015} (they both aim at highlighting non-gyrotropies in the 3D space)
and gives similar results, it will be only shown in the supporting material.

In addition to the analysis, the electron velocity distributions for some particular region are given, including the X-, D- and M-regions in order to reveal their characterizing signatures.

The paper is structured as follow. Section \ref{sec:simulation} gives further details on the simulation setup. Section \ref{sec:results} describes the most interesting results. 
Finally, the main conclusions are summarized in Section \ref{sec:conclusions}.

\section{Simulation Setup} \label{sec:simulation}

Results are shown for a set of 2.5D 
simulations performed with the fully kinetic massively parallel implicit moment method Particle-in-Cell 
code iPIC3D \citep{markidis2010, innocenti2016}. In 2.5D simulations all the vector quantities are three dimensional, 
but their spatial variation is assumed to be independent of the dawn-dusk ($Z$) direction.
A cartesian frame of reference is adopted, with the $X$ coordinate parallel to the initial current sheet (North-South direction in GSM), 
the $Y$ the direction parallel to the reversing $B$  
(Earth-Sun direction in GSM) and the $Z$
the direction to complete the set accordingly (dawn-dusk direction in GSM).
The simulated domain is 
a $40 \times 40\  d_i$ box with $2048 \times 2048$ cells, where $d_i = \frac{c}{\omega_{p,i}}$ is the ion skin 
depth referred to the magnetosheath conditions, to which all lengths in the code are normalized. 
The boundary conditions are periodic in all directions.
The temporal step is $\omega_{c,e} \cdot dt = 0.128$, where $\omega_{c,e}$ is the electron cyclo-frequency.
The ion-electron mass ratio is $\nicefrac{m_i}{m_e} = 256$. The plasma temperature across the layer is kept constant with an ion-electron temperature ratio $T_i = 2 T_e$.
The initial electron velocity is $\nicefrac{V_{th,e}}{c} = 0.1$, and $\nicefrac{c}{V_A} = 113.6$, where $V_A$ is the Alfv\'en speed. 
The initial ion drift is neglected, and the initial current density is fully carried by electrons, as done in the literature \citep{pritchett2007}.
With the parameter here considered, we simulate a plasma with plasma beta $\beta_{sh} \sim 1.2$ in the magnetosheath side and $\beta_{sp} \sim 0.05$ in the magnetosphere side. These values 
are compatible with those commonly observed in these regions, e.g., $\beta_{sh} \sim 2.4 \text{ and } \beta_{sp} \sim 0.27$ \citep{cassak2016}.
Two current sheets are configured at $y_1 = \frac{1}{4} L_y = 10\ d_i$ and $y_2 = \frac{3}{4} L_y = 30\ d_i$, 
with a current sheet half-thickness $L = 0.5\ d_i$. The magnetic field and density profiles across the current sheets are shown and described in the supporting material.
We simulate two different current sheets to compare two different reconnection mechanisms.

The upper layer (i.e. the one centered at $y_2 = 30\ \unit{d_i}$) is described by a continuous hyperbolic function \citep[e.g.][]{quest1981,pritchett2008,pritchett2009}. 
An initial out-of-plane current density is initialized according to $\nabla \times \nicefrac{\mathbf{B}}{\mu_0} = \mathbf{J}$. 
An initial perturbation identical to that used in \citet{lapenta2010} is set in the middle of the layer to produce a single X-point.
In contrast, the lower layer (i.e. the one centered at $y = 10\ \unit{d_i}$) is configured as a pure tangential discontinuity under an extremely steep gradient, 
with the same asymmetric profiles as the upper layer. No current density is initially set, 
causing this layer to be intrinsically highly unstable and suitable to gain more insights into physics of island coalescence  in asymmetric reconnection. 
No initial perturbation is added to this layer. 
A strong current density is naturally formed 
during the very first stages of the simulation to counteract the initial imbalance.
The same simulation is carried out with and without an initial guide field  
$B_g = B_{0,x}$.

\section{Results} \label{sec:results}

\begin{figure*}
 \centering
 \noindent\includegraphics[scale=.7]{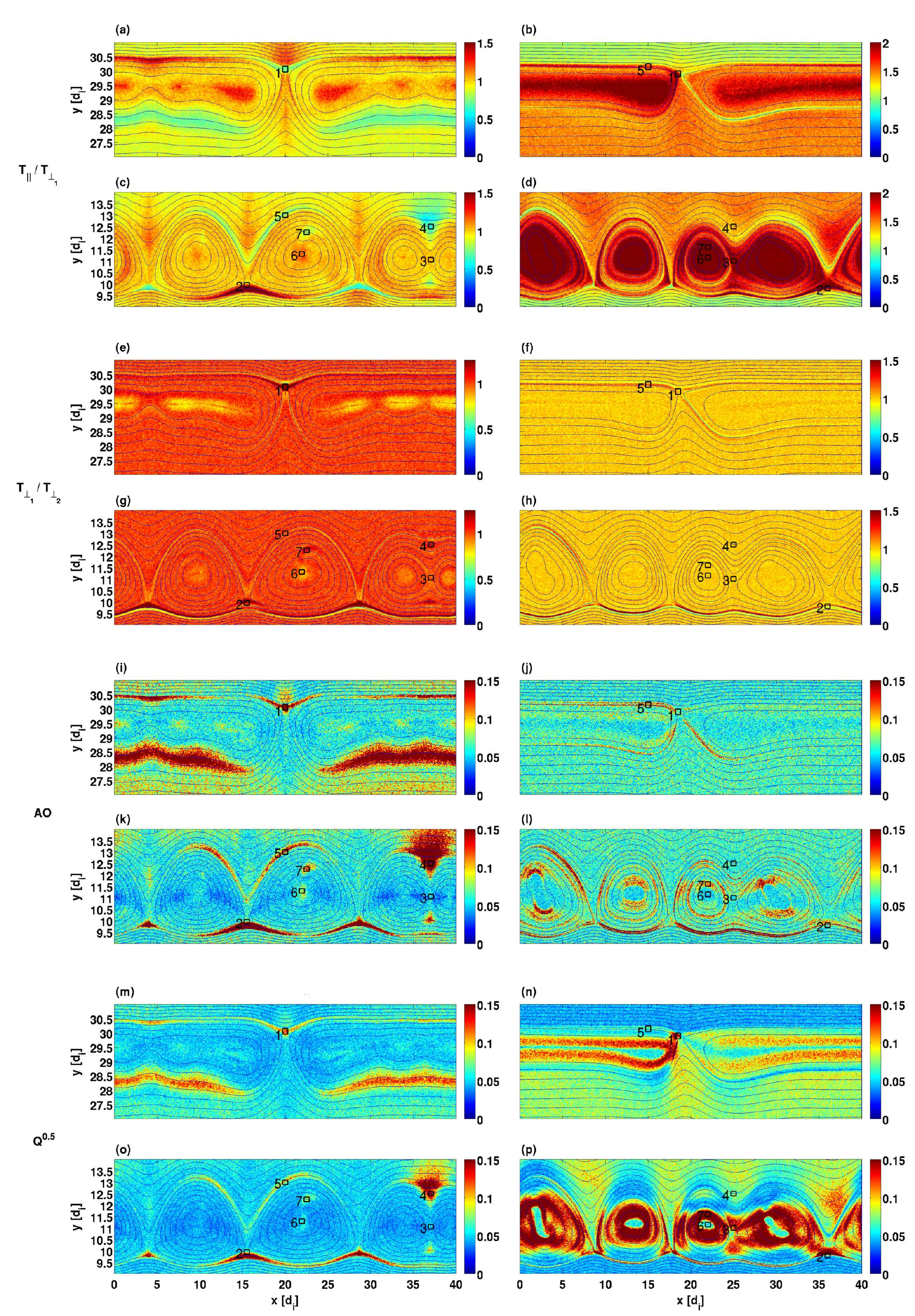}
 \caption{Plot of $\nicefrac{T_{||}}{T_{\perp_1}}$, $\nicefrac{T_{\perp_1}}{T_{\perp_2}}$, $A\O$ and $\sqrt{Q}$ for the two current sheets at $t \sim 21\ \unit{\omega_{ci}^{-1}}$, for the case with no guide field (left panels) 
 and with guide field (right panels). Black boxes indicate the domains considered for the phase-spaces.}
 \label{fig:superFig}
 \end{figure*}

Figure \ref{fig:superFig} shows four different quantities in the two current sheets at $\omega_{c,i} \cdot t \sim 21$, 
for the case with no guide field (lefthand panels) and the case with guide field (righthand panels). 

As already pointed out in \citet{cazzola2015}, at this time step the upper layer features a typical asymmetric 
reconnection site
\citep[e.g.][]{pritchett2007,pritchett2008,cassak2007,swisdak2003}. 
Meanwhile, the lower layer has rapidly evolved in the formation and growth of several magnetic islands, which in time progressively coalesce. 

A first analysis is carried out in a frame of reference based upon the local magnetic field direction, as suggested in 
\citet{goldman2015}. The parallel and perpendicular directions are identified as follows

\begin{eqnarray} 
  \mathbf{\hat{e}_{||}}  &:&  \mathbf{B} \times \mathbf{\hat{e}_{||}} = 0 \nonumber \\
  \mathbf{\hat{e}_{\perp_1}}&=& \mathbf{B} \times \mathbf{\hat{e}_{z}}  \\
  \mathbf{\hat{e}_{\perp_2}}&=& \mathbf{B} \times \mathbf{\hat{e}_{\perp_1}} = - \mathbf{\hat{e}_{z}} B^2 + \mathbf{B} \left(  \mathbf{\hat{e}_{z}} \cdot \mathbf{B} \right)  \nonumber
\end{eqnarray} \label{eq:Baligned}

Panels (a)-(h) in Figure \ref{fig:superFig} display the departure from the initial electron isotropy and
gyrotropy by plotting the ratio between $\nicefrac{T_{||}}{T_{\perp_1}}$ and $\nicefrac{T_{\perp_1}}{T_{\perp_2}}$ ($T$ is the electron
temperature). As $\nicefrac{T_{||}}{T_{\perp_1}}$ and $\nicefrac{T_{||}}{T_{\perp_2}}$ are very similar, the latter is not plotted here.
The $\nicefrac{T_{||}}{T_{\perp_1}}$ and $\nicefrac{T_{\perp_1}}{T_{\perp_2}}$ ratios give a quick information on the anisotropy and agyrotropy based on the simulation
frame (notice the $z$-dependence in Eqs. \ref{eq:Baligned}).
Quantities $A\O$ (\citet{scudder2008}, panels \emph{i - l}) and $\sqrt{Q}$ (\citet{swisdak2015}, panels \emph{m - p}) are instead computed to be independent of the simulation frame.
As $A\O$ varies between $0 \text{ and } 2$, here the range is normalized to $\left[ 0, 1 \right]$ for a better comparison with the quantity 
$\sqrt{Q}$, which ranges $\left[ 0, 1 \right]$.

One can see that some regions are highlighted
in all the panels 
in the case without guide field, such as 
the upper separatrices in the single reconnection point, the lower separatrices in the lower current sheet and, to a lesser extent, the reconnection exhausts
(i.e. at $x \sim 14 \text{ and } y \sim 29.5\ \unit{d_i}$) of the single X-point.
These regions then show both anisotropic and agyrotropic behavior.
Some other regions are instead only highlighted by $\nicefrac{T_{||}}{T_{\perp_1}}$, $A\O{}$ and $\sqrt{Q}$ and not seen in  $\nicefrac{T_{\perp_1}}{T_{\perp_2}}$, including: 
(1) the separatrices bordering the weaker field side, (2) the inflow region from the stronger field side (i.e. $x \sim 20 \text{, } y \sim 30.5\ \unit{d_i}$), and (3) the particular outflow observed 
between the islands near $x \sim 37\ \unit{d_i}$ in the lower current sheet.
The reason is that $\mathbf{\hat{e}}_{\perp_1}$ is, by definition, in the simulation plane, but $\mathbf{\hat{e}}_{\perp_2}$ is perpendicular to both $\mathbf{\hat{e}}_{\perp_1}$ and $\mathbf{B}$.
Hence, when $\mathbf{B}$ is mostly in the plane (as it preferentially happens in the case without guide field as opposed to the case with guide field), 
the perpendicular to $\mathbf{B}$ plane is nearly perpendicular to the simulation plane also.
 $\nicefrac{T_{\perp_1}}{T_{\perp_2}}$ shows the projection of temperature agyrotropy in the simulation
 plane. Thus, when $B_g = 0$, 
$\nicefrac{T_{\perp_1}}{T_{\perp_2}}$ captures some, but not all, of the agyrotropic regions.
$A\O$ and $\sqrt{Q}$ do a better job. Notice also that $A\O$ and $\sqrt{Q}$ are quite similar in the case without guide field.

Remarkable differences are identified in the case with guide field. 
The quantities $A\O{}$, $\sqrt{Q}$ and $\nicefrac{T_{\perp_1}}{T_{\perp_2}}$ show a different agyrotropic behavior in some peculiar regions. 
Examples are at separatrices, which are highlighted both in $\nicefrac{T_{\perp_1}}{T_{\perp_2}}$ and 
$A\O{}$ plots, but very weakly displayed in $\sqrt{Q}$, and the area within the magnetic islands, 
which are more powerfully marked by the quantity $\sqrt{Q}$ than $A\O{}$. Given the importance of separatrices in magnetic
reconnection \citep[e.g.][]{lapenta2014,Lapentabook}, this difference is of fundamental importance for satellite observations. 
One can notice that the areas within the islands highlighted by $\sqrt{Q}$ show a clear similarity with the anisotropy regions highlighted by $\nicefrac{T_{||}}{T_{\perp_1}}$ 
(panels \emph{b} and \emph{d}). This fact can be explained
because in the formula for $\sqrt{Q}$  the parallel pressure is accounted,
while $A\O{}$ is constructed considering only the plane perpendicular to the local magnetic field.
This is also the reason why the $A\O$ and $\nicefrac{T_{\perp_1}}{T_{\perp_2}}$ metrics show, in general, more similar results than $\sqrt{Q}$: in $\nicefrac{T_{\perp_1}}{T_{\perp_2}}$ and $A\O$ parallel pressure
is not considered.
Also, in the case with guide field, the $\hat{e}_{\perp_1}-\hat{e}_{\perp_2}$ plane considered in $\nicefrac{T_{\perp_1}}{T_{\perp_2}}$ and the perpendicular plane used in $A\O$ mostly superimpose.

Interesting is the analysis of the X-, D- and M-regions mentioned earlier, which can help distinguish the three regions during satellite observations.
In \cite{cazzola2015}, X- and D-regions were identified by comparing the $\nicefrac{T_{\perp_1}}{T_{||}}$ and $\nicefrac{T_{\perp_1}}{T_{\perp_2}}$ plots with the corresponding plots of the single X-point.
M-regions were instead 
identified with the reconnection exhaust flowing out
from an observed island merging.
Here, a similar analysis is performed for the case with guide field. 
By observing panels (d) and (h), we notice that all the reconnection sites resemble the traces
depicted in (b) and (f), with the only exception of the domain between $x = 35 \text{ and } 40\ \unit{d_i}$, which shows 
an opposite behavior. 
The latter is typical of a D-region, so it can be identified accordingly.
However, the upward-moving quadripolar structure seen in the D-regions in \citet{cazzola2015} and here visible in the case with no guide field 
(e.g. at $x \sim 15\ \unit{d_i}$ in panel \emph{g}) is no longer observed with guide field.
Moreover, the agyrotropic structures seen in the plasmoid centers are also absent.

\begin{figure*}
 \centering
 \noindent\includegraphics[scale=.55]{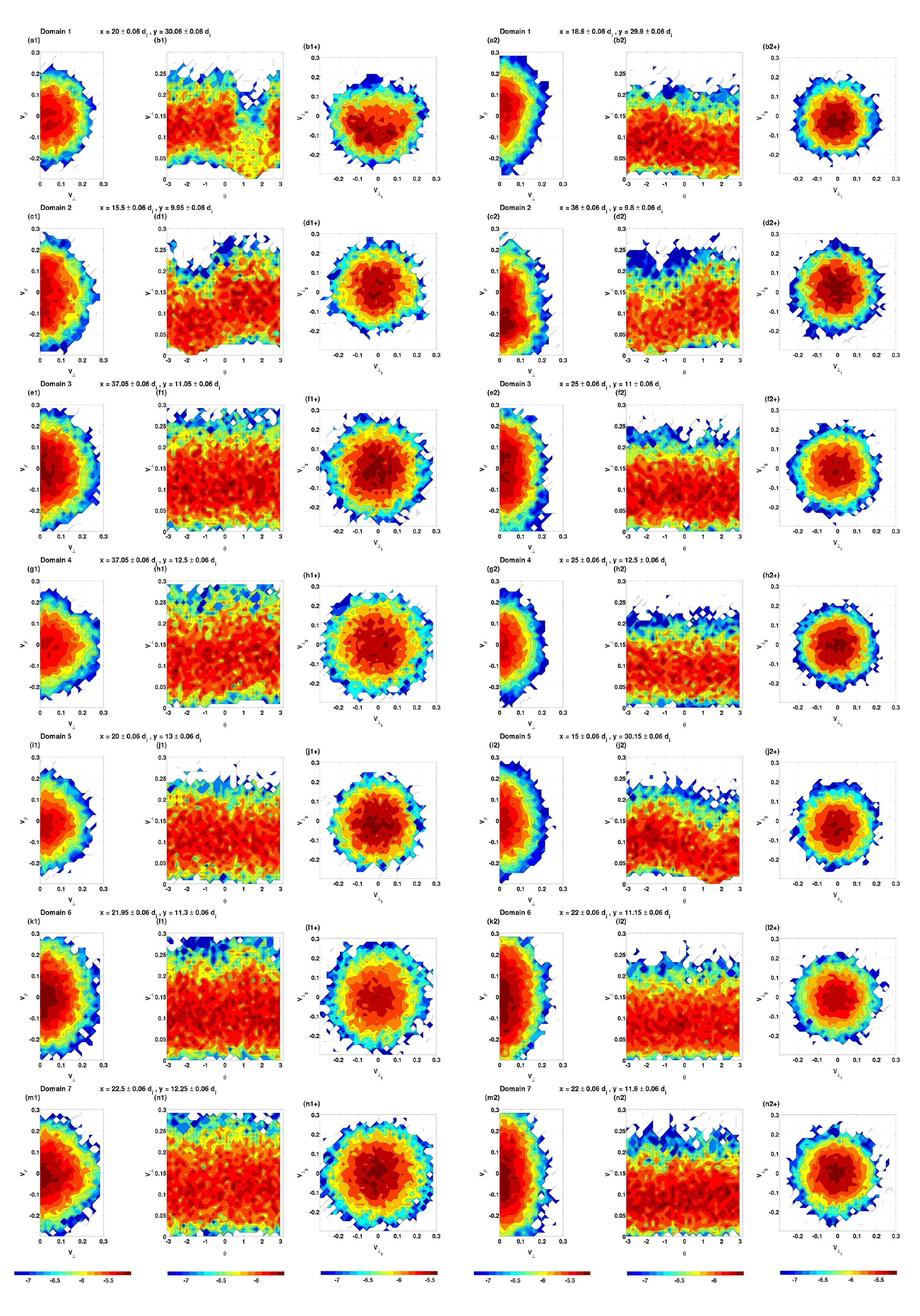}
 \caption{Set of electron velocity distributions for the domain pointed out with black boxes in figure \ref{fig:superFig}. Color scale indicates the number of particles, in logarithmic
 scale, over the infinitesimal volume-velocity domain. Panels ending with $1$ describe the case with no guide field, while those ending with $2$ the case with guide field. 
 Over the axis $V_{||}$, $V_{\perp} = \sqrt{V^2 - V_{||}}$ and
 $\theta$ as explained in the text. Finally, panels ending with $+$ represent phase-space in the $V_{\perp_1} - V_{\perp_2}$ plane. }
 \label{fig:vel_distr}
 \end{figure*}

In light of the upcoming satellite observations, we show the electron phase-space for the domains
marked with a black box in Figure \ref{fig:superFig}. 
Two different sets of electron phase-spaces are shown
in figure \ref{fig:vel_distr}, respectively, for the case without guide field (panels \emph{a1 - n1}) and with guide field (panels
\emph{a2 - n2}).
Each box (which is not to scale for a clearer representation), represents a physical domain of $0.12 \times 0.12\ \unit{d_i}$,
as a compromise between exact localization (which would require a smaller bin) and lack of noise in the representation.
Notice that
the same box number corresponds to comparable features in the case without and with guide field.
Domains $1$, $2$ and $3$ correspond to X-, D and M-regions for guide field values. Domain $4$ represents the situation at the outflow of the M-regions. Domain $5$ represents
the situation at the separatrices. Finally, Domains $6$ and $7$ study the situation, respectively, in the O-point and the inner region within a magnetic island.
Box numbers are replicated over all the plots for a better readability.
To represent velocity we introduce the coordinate system $\mathbf{V} = V_{||} \mathbf{\hat{b}} + V_{\perp} \mathbf{\hat{\Omega}}$, where $\mathbf{\hat{b}}$ is the magnetic field direction and
$\mathbf{\hat{\Omega}}$ is the direction in the
perpendicular plane. We plot the velocity distribution in the $V_{||} - V_{\perp}$ and $V_{\perp} - \theta$ plane, where 
$\theta \in \left[ - \pi, \pi \right]$ is the angle between the direction $\mathbf{\hat{e}_1}$ and $V_{\perp}$ in the plane normal to $\mathbf{\hat{b}}$.
Figure S2 in the supporting material gives a visual representation of how $\theta$ is calculated and why it spans the $\left[ - \pi, \pi \right]$ range.
Alongside these plots, we represent the phase-spaces in the same regions in the $V_{\perp_1} - V_{\perp_2}$ plane for a direct comparison (panels ending with $+$ in Figure \ref{fig:vel_distr}).
The color scale in Fig. \ref{fig:vel_distr} indicates the logarithm of the number of particles over the infinitesimal volume-velocity domain.

Regions of type X, Domains $1$, show a very remarkable agyrotropy for both the case with no guide field (panel \emph{b1}),
and, less marked, with guide field (panel \emph{b2}), as expected \citep{hesse2016,chen2016}. 
Additionally, in the case without guide field we notice the presence of a crescent-shape velocity distribution in the $V_{\perp_1} - V_{\perp_2}$ plane (panel \emph{b1+}). The same is not seen clearly
in the case
with guide field (panel \emph{b2+}), although expected \citep{hesse2016,chen2016,burch2016}. This effect is probably due to the presence of a relatively strong guide field, which tends to dampening the particles agyrotropic behavior, 
by smearing out any possible crescent outcome. The latter also explains the lower agyrotropic rate observed in panel (b2) compared to the case without guide field in panel (b1).

Finally, the case with guide field shows a relevant particle anisotropy (panel \emph{a2}), noticeable also from the $\nicefrac{T_{||}}{T_{\perp_1}}$ plot.
In the case with guide field, only the metric $\sqrt{Q}$ shows a relevant agyrotropy, mostly extended from the left outflow, whereas the trace in $A\O{}$ results very moderate.
We explain this effect with the particle anisotropy in the computation of $\sqrt{Q}$.

The situation in the D-regions, Domains 2, are similar with and without guide field (panels \emph{c1 - c2} and \emph{d1 - d2}). The flat-top velocity distribution typical of these regions \citep{cazzola2015} appears 
even more remarked in the case with 
guide field, associated with a strong anisotropy. The same agyrotropic features are detected in $A\O$ and $\sqrt{Q}$.

Interesting is the situation in the 
M-regions, i.e. Domains $3$. 
The velocity distributions in panels (f1) and (f2) do not show any relevant agyrotropic features in neither case. The 
algorithms
$\nicefrac{T_{\perp_1}}{T_{\perp_2}}$, $A\O{}$ and $\sqrt{Q}$ for the case
with no guide field also do not show any agyrotropy. However, the case with guide field is slightly different. While a null agyrotropy value is predicted by algorithms $\nicefrac{T_{\perp_1}}{T_{\perp_2}}$ and $A\O{}$, quantity $\sqrt{Q}$ 
instead indicates that a certain agyrotropy is
present close to the merging point.
We believe that this effect can be explained by
the presence of a strong parallel component in the region, as confirmed by $\nicefrac{T_{||}}{T_{\perp_1}}$.
$\sqrt{Q}$ shows that the electron distribution is not isotropic. With the help of $A\O$ and the phase-spaces, we understand that the lack of isotropy is mostly driven by a strong anisotropy,
which however does not exclude a moderate agyrotropy be present.
Concerning the M-regions, we analyze their vertical reconnection outflow in Domains $4$. In the case without guide field we observe all the quantities to highlight the presence of a strong agyrotropy, 
also confirmed by the corresponding phase-space (panel \emph{h1}). In the case with guide field, a clear reconnection outflow is not seen, stressing the atypical behavior being
held during the island merging in presence of a strong guide field.

We formerly mentioned that the main difference between $A\O$ and $\sqrt{Q}$ lies in the case with guide field at the separatrices and within the islands.
The separatrix is studied in Domain $5$, while Domains $6$ and $7$ focus on the situation within the magnetic islands.
From panels (j1) and (j2) in Figure \ref{fig:vel_distr}, we observe that, in the case without guide field, the separatrix shows a very moderate agyrotropy, similar to what pointed out by $\sqrt{Q}$. 
In the case with guide field, the agyrotropy is much better highlighted in panel (j2), which confirms what represented in $\nicefrac{T_{\perp_1}}{T_{\perp_2}}$ and $A\O$. Conversely, the quantity $\sqrt{Q}$
shows a much weaker agyrotropic signature in this region.
Finally, the situation within the islands is analyzed in Domains $6$ and $7$. Domain $6$ gives some insight into the island center, i.e. the O-point. In the case without guide field, signatures of
agyrotropy are detected by $\nicefrac{T_{\perp_1}}{T_{\perp_2}}$ and $A\O$, and less remarked
in $\sqrt{Q}$. The phase-space analysis confirms the presence of agyrotropy mainly shown for mid-energy electrons (greenish band).
Instead, in the case with guide field no agyrotropy is highlighted by either quantities, nor is by the corresponding phase-space (panel \emph{l2}).
This indicates that O-points in presence of a strong guide field show a different behavior compared to the traditional case without guide field.
Finally, Domain $7$ analyses the agyrotropic \emph{patch} visible in the case without guide field and already pointed out in \cite{cazzola2015}. In the case with guide field, the same Domain
gives information on the inner agyrotropic structure predominantly highlighted in $\sqrt{Q}$. From panels (n1) and (n2) we observe that an agyrotropic signature is present in the case without guide field, mostly confined
for mid-energy electrons (yellowish band), while the case with guide field does not show any clear agyrotropy, showing however a remarkable anisotropy.

\section{Conclusions} \label{sec:conclusions}

This work presents a systematic comparison of the electron agyrotropic behavior from PIC simulations of asymmetric magnetic reconnection during rapid island coalescence, 
with particular focus on the X-, D- and M-regions identified in \cite{cazzola2015}. Cases with and without guide field have been addressed.
Three detection algorithms for highlighting agyrotropy have been compared: the ratio between the perpendicular temperature components, the method proposed 
in \cite{scudder2008} and that in \cite{swisdak2015} (Fig. \ref{fig:superFig}).
Additionally, the ratio between the parallel and perpendicular components ($\nicefrac{T_{||}}{T_{\perp_1}}$) is used to highlight anisotropic regions.
Different regions have been analysed in terms of the electron velocity phase-space for a helpful comparison with observational data, including the X-, D- and M-regions pointed out in \cite{cazzola2015}
as well as other relevant regions.
A new representation method is adopted here
to better represent the relation between the velocity perpendicular components  $V_{\perp_1}$ and $V_{\perp_2}$. 
The phase-space in the same regions 
on the $V_{\perp_1} - V_{\perp_2}$ plane are also plotted
in Figure \ref{fig:vel_distr}. Unlike the $V_{\perp} - \theta$ representation, the latter seems less suited to highlighting agyrotropy features, 
except for the case in the X-regions, where important features
are particularly detected.

Below we provide a summary description of the main findings for each region analysed, as well as a brief comment on the performance of the different algorithms compared. 
Additionally, Table \ref{tab:sumTable} gives a wider and quicker summary 
of the features.

\paragraph{Methodology Remarks}
We observe that the ratios $\nicefrac{T_{||}}{T_{\perp_1}}$ and $\nicefrac{T_{\perp_1}}{T_{\perp_2}}$ give 
a quick and reliable initial insight into the anisotropy and agyrotropy.
However, these algorithms are simulation-frame dependent, unlike those from such as \citet{scudder2008} 
and \citet{swisdak2015}.

A noticeable discrepancy is detected between the two metrics $A\O$ and $\sqrt{Q}$ in some regions in the case with guide field. 
We observe that the detection of $\nicefrac{T_{\perp_1}}{T_{\perp_2}}$ is closer to $A\O$ than $\sqrt{Q}$.
We interpret this difference as due to a strong relevance of the parallel component. 
The parallel component is not included in the computation of $A\O$,
while it enters the calculation of $\sqrt{Q}$.
Since $\nicefrac{T_{\perp_1}}{T_{\perp_2}}$ does not consider the parallel component also, $\nicefrac{T_{\perp_1}}{T_{\perp_2}}$ and $A\O$ are tendentially similar.
It is interesting to comment on how similar $\nicefrac{T_{\perp_1}}{T_{\perp_2}}$ and $A\O$ plots are in the case with and without guide field.
In the case with guide field, 
the presence of a relevant parallel component in the out-of-plane
direction makes the plane perpendicular to the magnetic field nearly parallel to the simulation plane. This fact leads the representation of $\nicefrac{T_{\perp_1}}{T_{\perp_2}}$ to be particularly similar to $A\O$.
In the case without guide field, the plane perpendicular to the local magnetic field can take different directions with respect to the simulation plane.
Hence, $\nicefrac{T_{\perp_1}}{T_{\perp_2}}$ and $A\O$ and $A\O$ are more different in the case without guide field.

We now focus on specific regions.

\paragraph{Separatrices}
The agyrotropy in the separatrices is differently represented in the case with guide field. 
Separatrices are weakly highlighted in $\sqrt{Q}$ compared to $A\O{}$.
This is probably due to the parallel component accounted in the computation of $\sqrt{Q}$.
The velocity distributions analysis confirmed the presence of a clear agyrotropic behavior along the separatrices  in Domain $5$ for the case with guide field 
(panels \emph{j2} in 
Figure \ref{fig:vel_distr}).
However, the separatrix in Domain $5$ in the case without guide field shows a very moderate agyrotropy (panel \emph{j1}), in line with the prediction of $\sqrt{Q}$.

\paragraph{X-Regions}
These regions are present in both the cases with and without guide field. 
A relevant agyrotropy is observed in both the cases (panels \emph{b1} and \emph{b2}).
The strong anisotropy drives a detection of non gyrotropy from the $\sqrt{Q}$ metric. $A\O$ does not shown agyrotropy (Fig. \ref{fig:superFig}).
Interestingly, the analysis of the phase-space on the $V_{\perp1} - V_{\perp_2}$ plane shows a crescent shape in the case without field.
However, the same is not seen in the case with guide field, as instead expected \citep{hesse2016,chen2016}.

\paragraph{D-Regions}
Signatures of D-regions are observed also in the case with guide field.
The agyrotropy in these regions is well represented by all the models. 
The velocity distributions additionally reveal agyrotropy in the cases with and without guide field (panel \emph{d1} and \emph{d2}). 
The first case shows
a strong anisotropy and the typical flat-top velocity distribution pointed out in \cite{cazzola2015} (Fig. \ref{fig:vel_distr}, panels \emph{c1} and \emph{c2}). 

\paragraph{M-Regions and Outflow}
M-regions are found in the cases with and without guide field.
While the only noticeable agyrotropy is revealed by $\sqrt{Q}$ for the case with guide field, velocity distributions 
show no presence of agyrotropy in either case.
A remarked anisotropy is observed in these regions for the case with guide field.
We attribute the signature in $\sqrt{Q}$ to this latter.
Domain $4$ has been considered to analyse the reconnection outflow from the M-regions. In the case without guide field, a sharp agyrotropy trace is highlighted by all the quantities, and further confirmed
by the corresponding phase-space (panel \emph{h1}). The same Domain with guide field instead does not show any agyrotropy signature, neither from the detection algorithms nor from the phase-space.
\paragraph{Magnetic Islands}
The agyrotropic behavior within magnetic islands has been studied in Domains $6$ and $7$.
In particular, Domains $6$ represent the typical O-point.
In the case without guide field, $\nicefrac{T_{\perp_1}}{T_{\perp_2}}$, $A\O$ and $\sqrt{Q}$ all show an agyrotropic signature in the centre of the islands, even though with a different degree.
This agyrotropy is confirmed by the velocity distribution in panel (l1). 
The same Domain with guide field instead does not show any relevant agyrotropy, nor it does in the phase-space. This effect suggests that the O-points in the case with strong guide fields tend to
behave differently than when a guide field is absent.
Concerning the case with guide field, we observe important differences between $A\O$ and $\sqrt{Q}$ within the islands: while $A\O$ detects a moderate agyrotropy only around specific closed magnetic field lines, 
$\sqrt{Q}$ shows a more extended agyrotropic region within the islands (panels \emph{l} and \emph{p} in Fig. \ref{fig:superFig}).
Domains $7$ intends to analyse this situation.  
In the case without guide field, it represents the situation at the agyrotropic \emph{patch} already observed in \cite{cazzola2015}. The presence of agyrotropy is confirmed by all the algorithms, as well as
the corresponding phase-space (panel \emph{n1}).
In the case with guide field, the corresponding
phase-space shows that no relevant agyrotropy is present in this region.


%
%
%
%
%
%
%

\section*{Acknowledgments}
 
The present work is supported by the NASA MMS Grant
NNX08AO84G. Additional support for the KULeuven
team is provided by the European Commission DEEP-ER project, by the Onderzoekfonds KU Leuven (Research Fund KU Leuven) and 
by the Interuniversity Attraction Poles Programme of the Belgian Science Policy Office (IAP P7/08 CHARM). 
M.E.I. is funded by the FWO (Fonds Wetenschappelijk Onderzoek – Vlaanderen) postdoctoral fellowship reference 12O5215N. 
The simulations were conducted on the Pleiades supercomputer of the NASA Advanced Supercomputing Division (NAS), on the Discover supercomputer of the NASA Center for Climate Simulation (NCCS), 
on the computational resources provided by the PRACE Tier-0 framework and 
on the Flemish Supercomputing Center (VSC-VIC3).
The data produced by
the simulations are stored in HDF5 format on the NASA-NAS data servers.

\begin{sidewaystable}
\caption{Comparison Table among the different detection algorithms, including the velocity distribution analysis. Quantities $\nicefrac{T_{||}}{T_{\perp_1}}$, $\nicefrac{T_{\perp_1}}{T_{\perp_2}}$, 
$A\O$ and $\sqrt{Q}$ are described quantitatively, the distributions are described qualitatively as \emph{Not-detected, Low, Medium and Remarked.} 
}
\centering
\resizebox{\textwidth}{!}{\begin{tabular}{l | c | c | c | c | c | c}
\hline
\multicolumn{7}{c}{\textbf{Without Guide Field}} \\
  \hline
\textbf{Region}  & $\mathbf{\nicefrac{T_{||}}{T_{\perp_1}}}$ & $\mathbf{\nicefrac{T_{\perp_1}}{T_{\perp_2}}}$ & $\mathbf{A\O}$  & $\mathbf{\sqrt{Q}}$  & $\mathbf{V_{||} - V_{\perp}}$ & $\mathbf{\theta - V_{\perp}}$  \\
\hline

Method Features & \multicolumn{2}{|c|}{\begin{tabular}{@{}c@{}}Frame-dependent \\ Useful for quick assessments \\ Qualitative results \\ Results similar to both $A\O$ and $\sqrt{Q}$ \end{tabular}} 
& \multicolumn{2}{|c|}{\begin{tabular}{@{}c@{}}Quantitative results \\ $A\O$ computed in the plane, $\sqrt{Q}$ computed 3D \\ Strong agreement between the methods  \end{tabular}} & 
\multicolumn{2}{|c}{\begin{tabular}{@{}c@{}}Direct Comparison with Satellite observations \\ Direct and Clear Assessment of anisotropy and agyrotropy \end{tabular}} \\ \hline

 Region X (Domain 1)  & $T_{||} < T_{\perp}$ & $T_{\perp_1} > T_{\perp_2}$ & $15 \%$ & $10 - 15 \%$ & low & \begin{tabular}{@{}c@{}} Remarked \\ Crescent-shape distribution on $V_{\perp_1} - V_{\perp_2}$ plane \end{tabular} \\ \hline
 Region D (Domain 2)  & $T_{||} > T_{\perp}$ & $T_{\perp_1} > T_{\perp_2}$  & $15 \%$  & $15 \%$ & \begin{tabular}{@{}c@{}} Remarked \\ double beam \\ flat top distribution \end{tabular} & Remarked \\ \hline
 Region M (Domain 3)  &  $T_{||} \sim T_{\perp}$ & $T_{\perp_1} \sim T_{\perp_2}$ & Not-detected & Not-detected  & Low & Not-Detected \\ \hline
 Region M - Outflow (Domain 4)  &  $T_{||} < T_{\perp}$ & $T_{\perp_1} \sim T_{\perp_2}$ & $15 \%$ & $15 \%$ & Relevant for slow electrons & Medium \\  \hline
 Separatrix (Domain 5) & $T_{||} < T_{\perp}$ &  $T_{\perp_1} \sim T_{\perp_2}$ &  $15 \%$ & $10 - 15 \%$ & Relevant for slow electrons & \begin{tabular}{@{}c@{}}Low \end{tabular} \\ \hline
 Magnetic Island Center - O-point (Domain 6) & $T_{||} > T_{\perp}$ & $T_{\perp_1} < T_{\perp_2}$  & $10 \%$ & $ 5 - 10 \%$ & Low & Remarked for mid-energy electrons \\ \hline
  Magnetic Island - Inner \emph{Patch} (Domain 7) & $T_{||} < T_{\perp}$ & $T_{\perp_1} \sim T_{\perp_2}$  & $15 \%$ & $ 5 - 10 \%$ & Low & Medium for mid-energy electrons \\ 
\hline
\multicolumn{7}{c}{\textbf{With Guide Field}} \\
  \hline
\textbf{Region}  & $\mathbf{\nicefrac{T_{||}}{T_{\perp_1}}}$ & $\mathbf{\nicefrac{T_{\perp_1}}{T_{\perp_2}}}$ & $\mathbf{A\O}$  & $\mathbf{\sqrt{Q}}$  & $\mathbf{V_{||} - V_{\perp}}$ & $\mathbf{\theta - V_{\perp}}$  \\
\hline

Method Features & \multicolumn{2}{|c|}{\begin{tabular}{@{}c@{}}Frame-dependent \\ Useful for quick assessments \\ Qualitative results \\ Results similar to both $A\O$ and $\sqrt{Q}$ \end{tabular}} 
& \multicolumn{2}{|c|}{\begin{tabular}{@{}c@{}}Quantitative results \\ $A\O$ computed in the plane, $\sqrt{Q}$ computed 3D \\ \emph{Relevant disagreement} between the methods in some regions, such as separatrices \end{tabular}} & 
\multicolumn{2}{|c}{\begin{tabular}{@{}c@{}}Direct Comparison with Satellite observations \\ Direct and Clear Assessment of anisotropy and agyrotropy \end{tabular}} \\ \hline

 Region X (Domain 1)  & $T_{||} > T_{\perp}$ & $T_{\perp_1} > T_{\perp_2}$  & $ 5 \%$ & $ 15 \%$ only on one side & Low & Remarked \\ \hline
 Region D (Domain 2)  & $T_{||} > T_{\perp}$ & $T_{\perp_1} > T_{\perp_2}$ & $15 \%$ & $ 15 \%$ & \begin{tabular}{@{}c@{}}Strong \\ double beam \\ flat top distribution \end{tabular} & Remarked \\ \hline
 Region M (Domain 3)  & $T_{||} > T_{\perp}$ & $T_{\perp_1} \sim T_{\perp_2}$  & Not-detected & $ 15 \% $ & Strong & Not-detected \\ \hline
 Region M - Outflow (Domain 4) & $T_{||} \sim T_{\perp}$ & $T_{\perp_1} \sim T_{\perp_2}$  & Not-detected &  Not-detected & Low & Not-Detected \\ \hline
  Separatrix (Domain 5)  & $T_{||} > T_{\perp}$  & $T_{\perp_1} > T_{\perp_2}$ & $ 15 \% $ & Not-detected & Relevant for slow electrons & Medium \\ \hline
  Magnetic Island Center - O-point (Domain 6)  & $T_{||} > T_{\perp}$  & $T_{\perp_1} \sim T_{\perp_2}$ & Not-detected & Not-detected & Remarked & Not-detected \\ \hline
 Magnetic Island - Inner Closed Lines (Domain 7) & $T_{||} > T_{\perp}$  & $T_{\perp_1} \sim T_{\perp_2}$ &  $ \sim 10 \% $ & $ 15 \% $ & Remarked & Not-detected \\
 \hline
\end{tabular}} \label{tab:sumTable}
\end{sidewaystable}

\end{document}


\maketitle
%
%

\noindent\textbf{Contents of this file}
\begin{enumerate}
\item Figure S1
\item Figure S2
\item Figure S3

\end{enumerate}

\noindent\textbf{Introduction}

\begin{figure}
\centering
 \noindent\includegraphics[scale=.6]{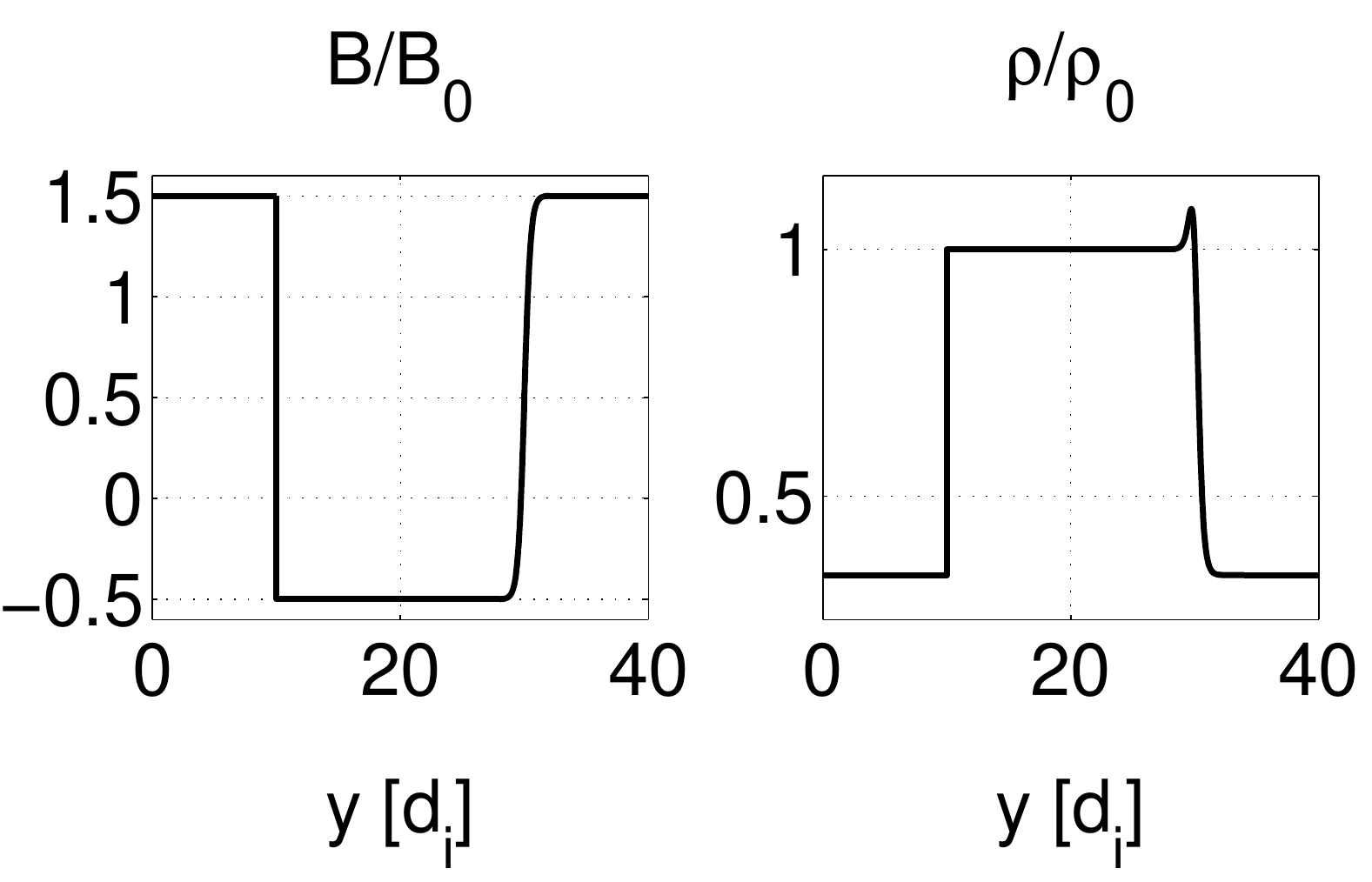}
  \caption{Initial profiles of magnetic field (B) and density ($\rho$) adopted. Notice how the periodicity is being respected at the boundaries.}
 \label{fig:profiles}
 \end{figure}

First of all, we report  the  magnetic field and density initial profiles adopted for the analysis

\begin{subequations} \label{eq:Beq}
\begin{empheq}[left={ B_{x} \left( y \right) = }\empheqlbrace]{align}
  & \frac{3}{2} B_0 & \text{$y \le \frac{L_y}{4}$} \\ 
  &  B_0 \left[ \tanh \left(\frac{y - y_2}{\lambda} \right) + R \right]  & \text{$y > \frac{L_y}{4}$} \label{eq:Beqb}
\end{empheq}
\end{subequations}

 \begin{subequations} \label{eq:neq}
\begin{empheq}[left={n\left( y \right) = }\empheqlbrace]{align}
 & n_0 \left(1 - 2 \alpha \right) & \text{$y \le \frac{L_y}{4}$} \\
\begin{split} & n_0 \left[ 1  - \alpha \tanh \left(\frac{y - y_2}{\lambda} \right) \right. \\
      &- \left. \alpha \tanh^2 \left( \frac{y - y_2}{\lambda} \right) \right] \end{split} & \text{$y > \frac{L_y}{4}$} \label{eq:neqb}
\end{empheq}
\end{subequations}
where $y_2 = \frac{3}{4} Ly$, $\alpha = 0.33$ and $R = 0.5$, which lead to the profiles in Fig. \ref{fig:profiles}.

\begin{figure}
\centering
 \noindent\includegraphics[scale=.6]{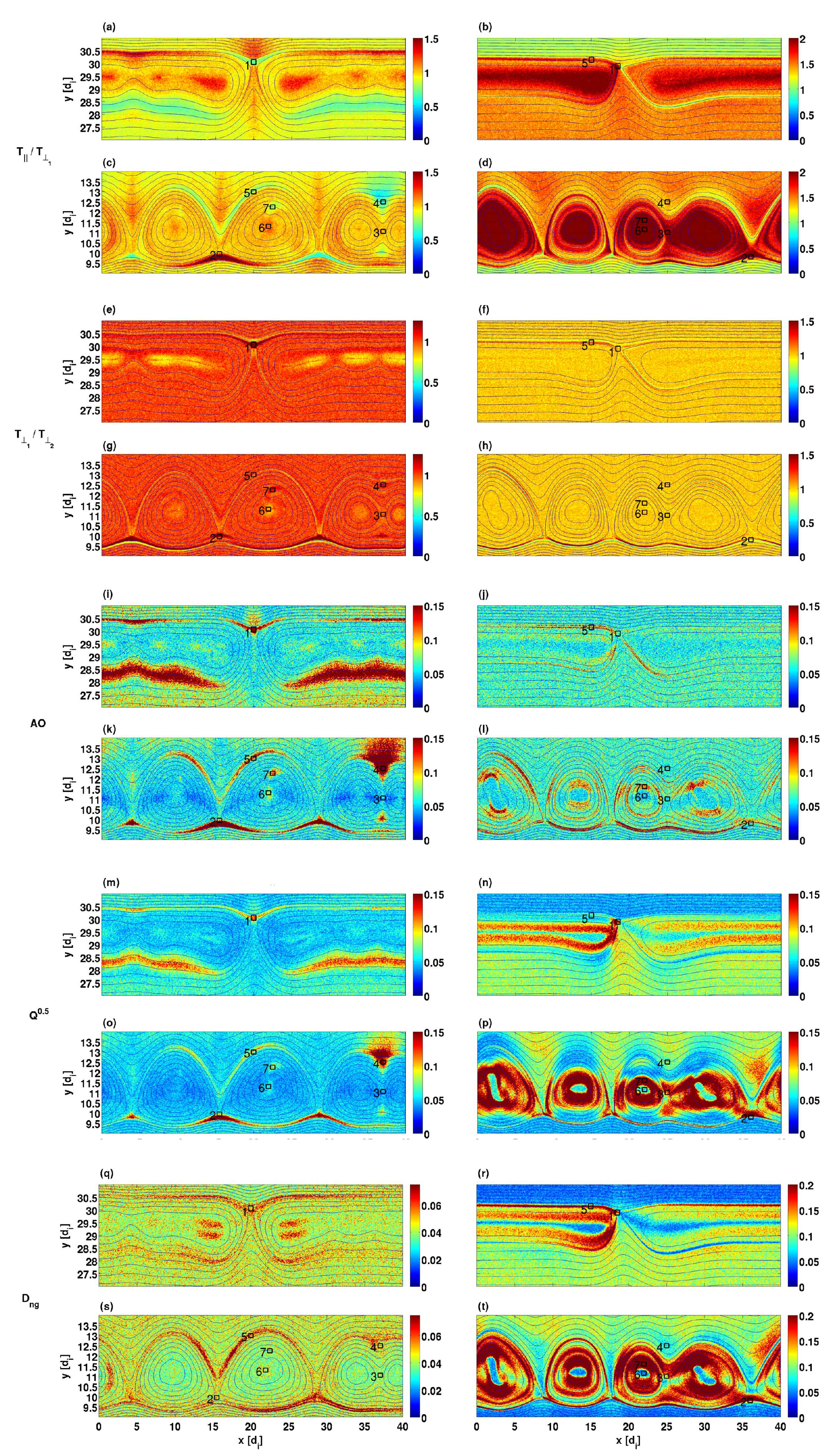}
  \caption{Plot of the agyrotropy detection algorithms proposed in 
 \citep{scudder2008} ($A\O$), \citep{swisdak2003} ($\sqrt{Q}$)  and \citet{aunai2013} ($D_{ng}$) for the two current sheets at $t \sim 21\ \unit{\omega_{ci}^{-1}}$, for the case with no guide field (left panels) 
 and with guide field (right panels). Black boxes indicate the domains considered for the phase-spaces shown in the manuscript. All the quantities are normalized to the respective maxima.
  Refer to the text for further information.}
 \label{fig:superFigAunai}
 \end{figure}

In this supporting material section we include a more complete comparison of agyrotropy detection algorithms. In particular, we add the algorithm
proposed in \citet{aunai2013}
(panels (q) through (t) in Fig. \ref{fig:superFigAunai}).
Figure \ref{fig:superFigAunai} shows the agyrotropic regions as computed from the algorithms proposed in \citet{scudder2008} ($A\O{}$), 
\citet{swisdak2015} ($\sqrt{Q}$) 
and \citet{aunai2013} ($D_{ng}$). 
The mathematical formulation of these three metrics is 
\begin{eqnarray} 
  A\O &=& 2 \frac{ \left| \lambda_3 - \lambda_2 \right| }{\lambda_3 + \lambda_2} \nonumber \\
  D_{ng} &=& \frac{\sqrt{ 8 \left( P_{xy}^2 + P_{xz}^2 + P_{yz}^2 \right) }}{P_{||} + 2 P_{\perp}} \\
  Q &=& 1 - \frac{4 I_2}{\left( I_1  - P_{||} \right) \left( I_1 + 3 P_{||} \right) }  \nonumber
\end{eqnarray} \label{eq:metrics}
where $\lambda_{2,3}$ are the non-trivial eigenvalues from the diagonalization of the pressure tensor constructed from distribution function 
perpendicular to the local magnetic field 
(see \cite{scudder2008} for further details), $I_1 = P_{xx} + P_{yy} + P_{zz}$ and $I_2 = P_{xx} P_{yy} + P_{xx} P_{zz} + P_{yy} P_{zz} - \left( P_{xy} P_{yx} + P_{xz} P_{zx} + P_{yz} P_{zy} \right)$, and 
$P_{ij}$ is the $\left( i,j \right)-th$ term of the pressure tensor.
All the quantities have been normalized to their maxima, with the maximum value  for $D_{ng}$ set to $\sqrt{8/3}$, as
pointed out in \citet{swisdak2015}. As done in the manuscript, the cases with  and without guide field (respectively, right panels and left panels) 
at the same time are compared ($\omega_{c,i} \cdot t \sim 21$).
as can be noticed, \citet{swisdak2015} and \citet{aunai2013}'s algorithms show similar results. These metrics both consider a 
tridimensional rendering, by including the parallel component of the pressure tensor in the assessment.
Some small differences are however still noticeable, 
for example the value of agyrotropy for 
different scale needed in the two cases: 
for the case without guide field $D_{ng}$ is around half of $\sqrt{Q}$, whereas the situation is reversed in presence of a guide field, where $D_{ng}$ is 
nearly double of $\sqrt{Q}$. 
This last fact highlights once more the importance of the parallel component in the pressure tensor, mostly 
in presence of a guide field, which strongly influence the parallel behavior of particles and needs to be taken into account.
Finally, another important difference is noticeable at the separatrices, where the quantity $D_{ng}$ is more marked than $\sqrt{Q}$. 

\begin{figure}
\centering
 \noindent\includegraphics[scale=.3]{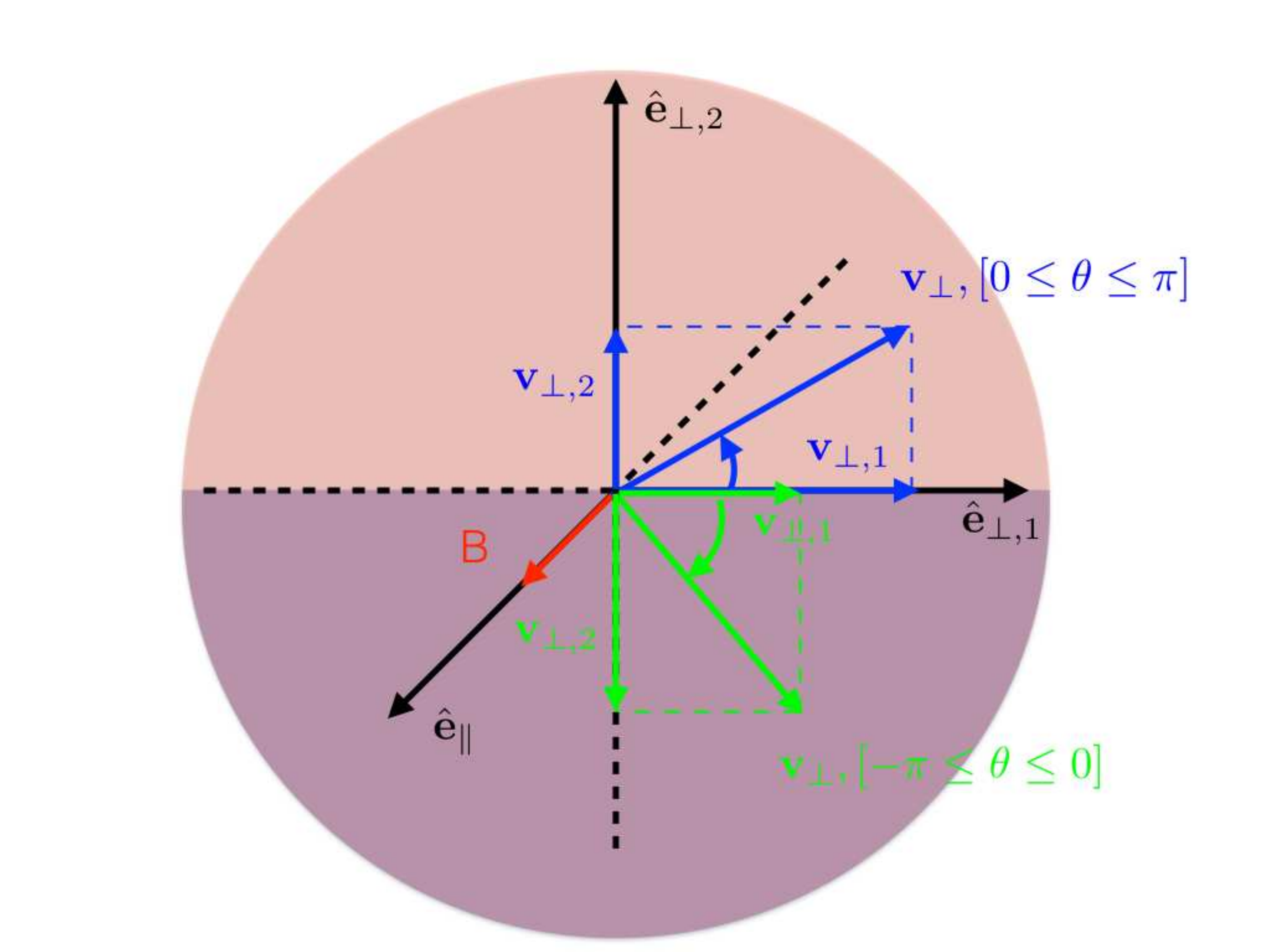}
  \caption{Visual representation of the coordinate system $V{||}, V_{\perp} \text{ and } \theta$ used for rendering agyrotropy and anisotropy in the phase-spaces in Figure 2 in the manuscript.}
 \label{fig:theta}
 \end{figure}

Additionally, Figure \ref{fig:theta} gives a geometrical representation of the coordinate system $\mathbf{V} = V_{||} \mathbf{\hat{b}} + V_{\perp} \mathbf{\hat{\Omega}}$ 
considered to highlight agyrotropy and anisotropy in the phase-spaces in Figure 2 of the manuscript.
In particular, $\mathbf{\hat{b}}$ is the direction parallel to the local magnetic field, $\mathbf{\hat{\Omega}}$ is the direction in the
perpendicular plane, $\theta \in \left[ - \pi, \pi \right]$ and $V_{\perp} = \sqrt{V^2 - V_{||}^2}$.

